\documentclass[letterpaper, 10 pt, conference]{ieeeconf}  

\IEEEoverridecommandlockouts                              

\usepackage{subcaption}
\DeclareCaptionFont{mysize}{\fontsize{8}{9.6}\selectfont{}}
\usepackage{adjustbox}
\usepackage{cite}

\usepackage{array}
\newcolumntype{C}[1]{>{\centering\arraybackslash}m{#1}}

\usepackage{multirow}
\usepackage{graphics} 
\usepackage{epsfig} 
\usepackage{times} 
\usepackage{amsmath} 
\usepackage{amssymb}  
\DeclareMathOperator*{\argmin}{argmin}
\title{\LARGE \bf
An Experimental Study on Relative and Absolute Pose Graph Fusion for Vehicle Localization 
}

\author{Anweshan Das and Gijs Dubbelman
\thanks{Anweshan Das and Gijs Dubbelman are with the Faculty of Electrical Engineering, Mobile Perception Systems group of SPS/VCA, Eindhoven University of Technology, 5612 AZ Eindhoven, The Netherlands.
        {\tt\small anweshan.das@tue.nl; g.dubbelman@tue.nl}}%
\thanks{This research has been sponsored by H2020 grant for the project INLANE.}%
    }

\begin{document}

\maketitle
\thispagestyle{empty}
\pagestyle{empty}


\begin{abstract}
	
	In this work, we research and evaluate multiple pose-graph fusion strategies for vehicle localization. We focus on fusing a single absolute localization system, i.e. automotive-grade Global Navigation Satellite System (GNSS) at 1 Hertz, with a single relative localization system, i.e. vehicle odometry at 25 Hertz. Our evaluation is based on 180 Km long vehicle trajectories that are recorded in highway, urban and rural areas, and that are accompanied with post-processed Real Time Kinematic GNSS as ground truth. The results exhibit a significant reduction in the error's standard deviation by 18\% but the bias in the error is unchanged, when compared to non-fused GNSS. We show that the underlying principle is the fact that errors in GNSS readings are highly correlated in time. This causes a bias that cannot be compensated for by using the relative localization information from the odometry, but it can reduce the standard deviation of the error.
\end{abstract}

\section{INTRODUCTION}

Autonomous driving technologies are evolving rapidly with the ultimate goal to develop a safe and reliable fully autonomous vehicle, i.e. SAE level 4 and eventually level 5 \cite{sae}. In this paper, our goal is to research and evaluate the performance and reliability of the graph-based fusion strategies with vehicle odometry and GNSS as input (GPS, GLONAS, Galileo are all GNSS systems). An accurate and robust localization system is the backbone of a fully autonomous vehicle. It is the basis for environment perception, path planning, and autonomous decision making by the vehicle.

 Generally, there are two different approaches to sensor fusion, namely filter-based and graph-based. Filter-based approaches, typically different variants of Bayes filters \cite{bayf,kalf,ekf,paf,rpaf}, are computationally less demanding and can easily meet real-time system constraints. Graph-based methods are designed to handle non-linear systems (solving non-linear optimization problems like SLAM \cite{gosl} \cite{gomsl} and bundle adjustment \cite{Ba}) and thus more accurate than filter-based methods \cite{whyfil}, but are computationally more demanding. However, with recent advancements in computational power and efficiency of processors, graph-based sensor fusion approaches are able to meet real-time system constraints. For a comparison of filter-based methods and graph-based methods, we refer to \cite{whyfil}.

\begin{figure}[]
	\centering
	\includegraphics[angle=0, scale=0.39,trim={0cm -.5cm 0cm -1cm},clip ]{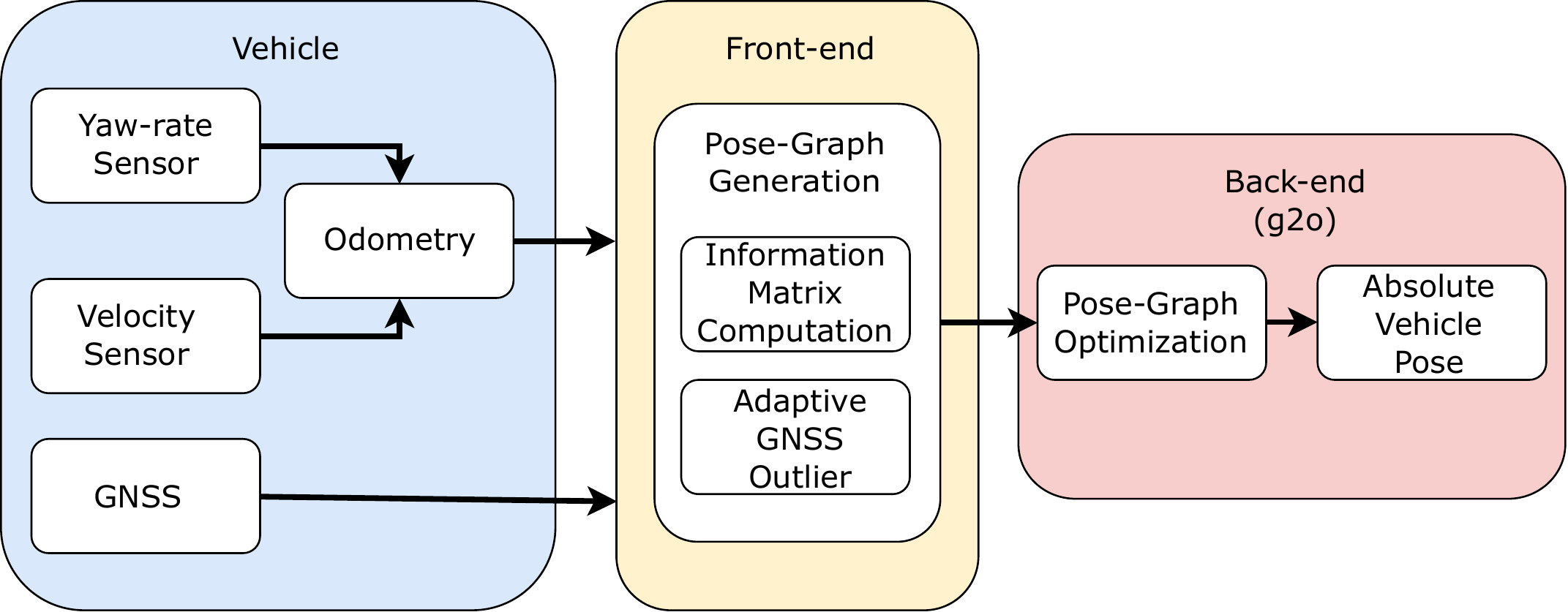}
	\caption{Graph optimization framework: The vehicle odometry is computed from yaw-rate and velocity sensors in the vehicle. The optimizable pose-graph is generated from vehicle odometry and GNSS measurements, and optimized using the g2o framework \cite{kuem1}. }
	\label{fig:framework}
\end{figure}

In our experiments, we use a low-cost state-of-the-art GNSS with Precise Point Positioning (PPP) \cite{pps}. It obtains accuracies of approximately 1 meter Circular Error Probable (\textit{CEP}), which is approx. equivalent to 1.48 standard deviation. The benefit of absolute GNSS positioning, compared to integrating relative positioning (vehicle odometry), is that the position errors are bounded, whereas integrating relative positioning will accumulate errors indefinitely (drift).

When fusing, relative position estimates do not contain direct information on the absolute position. Instead, they act as soft constraints between absolute poses. If the relative pose estimates would contain no errors (i.e. act as hard constraints) and if GNNSS readings would not be correlated in time, then absolute position errors could be reduced with the order of $ \sigma/\sqrt{n} $, with $ \sigma $ being the standard deviation of the GNSS error and $ n $ the number of GNSS readings. In reality, this model is highly naive because: 1) the relative pose estimates do contain errors and can only be used as soft constraints, and 2) the GNSS readings are highly correlated in time \cite{gpscorre}, reducing the statistical information of a single GNSS reading when fusion it with other localization sources. To gain more insight in the error reduction that one can obtain under realistic circumstances, this work performs an experimental study of several pose-graph fusion strategies and compares their output to post-processed RTK-GPS using an accurate 6 DoF IMU and corrections of a private RTK-base-station.

Our research extends the work by Merfels et al \cite{merf1}, where they have presented a graph-based framework to fuse multiple absolute and multiple relative localization systems in a plug-and-play manner. They generated sparse and computationally tractable pose-graph for online optimization from multiple sources. In their framework, the uncertainty of the GNSS readings were kept constant, hence the same for each GNSS reading. Evaluation of their approach was done using data recorded with their prototype vehicle with three absolute localization systems and one relative localization system for 16 Km driving distance.

We extend \cite{merf1} by taking into consideration GNSS dynamic uncertainties and by adding adaptive GNSS outlier rejection, as explained in Sec. \ref{sec: pose-graph-fusion}. In contrast to \cite{merf1}, we are fusing pose information from only one absolute (GNSS) and one relative (vehicle odometry) localization source. As there is no unique approach to project the vehicle localization task on basis of GNSS and vehicle odometry into a graph optimization framework, we put forward and evaluate three different approaches. These three approaches are detailed in Sec. \ref{sec:graph_struc_gen} and the overview of our entire framework is depicted in Fig. \ref{fig:framework}.
  
  To summarize, the main contributions of this paper are:
  \begin{itemize}
  	\item Development of a graph-based sensor fusion framework which generates sparse optimizable pose-graphs from vehicle odometry and GNSS observations, which extends \cite{merf1} by providing dynamic GNSS uncertainty modeling and adaptive GNSS outlier rejection.
  	\item Extensive analysis of the sensor fusion framework on large datasets covering more than 180 Km in different scenarios such as urban-canyons and highways. Thereby, providing fundamental insights in accuracy and precision of GNSS-odometry fusion.  
  \end{itemize}
Our experiments and results are provided in Sec. \ref{sec:exp_result} and our conclusion in Sec. \ref{sec:conclusion}. Furthermore, our data-sets are shared with the research community at \cite{dataset}.

\section{POSE GRAPH FUSION} \label{sec: pose-graph-fusion}

 The input data to the sensor fusion are pose measurements from vehicle odometry and GNSS receiver readings, represented as a pose-graph. The pose-graph consist of nodes, denoted by $ X $, which model the absolute vehicle pose by elements of $ SE(2) $, i.e. Euclidean motions in 2-D, and of edges denoted by $ Z $, which model the relative poses between nodes, also with elements of $ SE(2) $. Each measurement is accompanied with uncertainty expressed in the tangent space of $ SE(2) $ using an information matrix denoted with $ \Omega $. The edges always connect two nodes, i.e. the edge $ Z_{ij} $ denotes a relative pose that moves the node $ X_{i} $ onto $ X_{j} $. The error $ e_{ij} $ between the poses of the nodes $ X_{i} $ and $ X_{j} $ with respect to the measured relative pose $ Z_{ij} $ is computed with 
  \begin{equation}\label{eq:2}
 e_{ij}=\text{log}({Z_{ij}^{-1}}(X_{i}^{-1} X_{j}))
 \end{equation} 
 , where $ \text{log}() $ denotes the logarithmic map from $ SE(2) $ to its tangent space, i.e. $ e_{ij} $ is a three dimensional vector consisting of the angular and positional difference between $ X_{i} $ and $ X_{j} $. The goal of graph optimization is to minimize the following non-linear objective function
  \begin{equation}\label{eq:1}
 X^{*}=\argmin_{X} \sum\limits_{\langle i,j\rangle \in C } e_{ij}^{T}\Omega_{ij}e_{ij}
 \end{equation}
, where $ C $ represents the set of all index pairs for which measurements are available. This optimization task can be done with the usual non-linear solvers like Levenberg-Marqaurdt, Gauss-Newton, or Dogleg \cite{dogleg}. In our work, we use the Dogleg solver  contained in the g2o graph optimization framework developed by K\"{u}mmerle et.al.\cite{kuem1}.

\begin{figure}[]
	\centering
\begin{subfigure}[t]{.22\textwidth}%
		\centering
		\includegraphics[angle=-90, trim={8cm 10cm 7.5cm 13cm},scale=.6,clip]{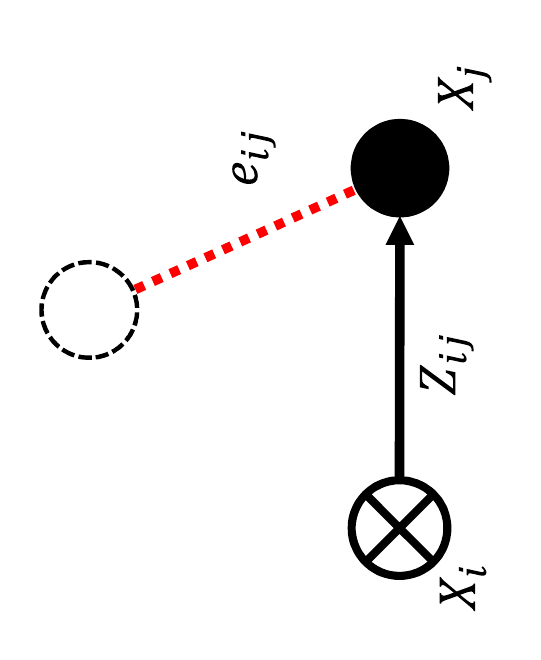}
	
		\caption{}
		\label{fig:graphdiagram2}
	\end{subfigure}
\begin{subfigure}[t]{.22\textwidth}%
\centering
\includegraphics[angle=-90, trim={8cm 10cm 7.5cm 14cm},scale=.6,clip]{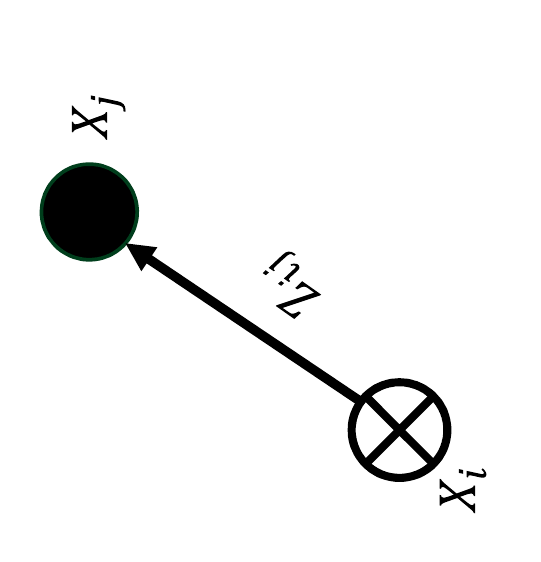}
\caption{}
\label{fig:graphdiagram3}
\end{subfigure}
\caption{A simple graph before (a) and after (b) optimization. The initial node $ X_{i} $ is kept fixed and before optimization the node $ X_j $ is at its initial position. The nodes $ X_{i} $ and $ X_{j} $ are connected by an edge (measurement) $ Z_{ij} $. The black dashed circle visualizes the measured position of  node $ X_j $ contained in the edge $ Z_{ij} $. The error vector $ e_{ij} $ before optimization is depicted as a red dashed line. After optimization this error is minimized by moving node $ X_j $ to the position according to the measurement contained in $ Z_{ij} $.}
\label{fig:graphdiagramall}
\end{figure}

In order to explain our three different graph structures, we use a graphical notation which is introduced next. Nodes or absolute poses in the graph are visualized using solid circles. Whenever a node is kept fixed, i.e. its pose is not optimized for, the circle contains a cross. The edges or relative poses are visualized using arrows. In order to better visualize the actual measurement contained in an edge and its error w.r.t. the absolute nodes, we visualize the measurement as a dashed circle and the error as a red dashed line. Fig. \ref{fig:graphdiagramall}, shows a simple graph before and after optimization. In this example the error is minimized by taking node $ X_{j} $ from its initial position to the position corresponding to the measurement contained in the edge $ Z_{ij} $. 

\begin{figure*}[]
	\centering
	\begin{subfigure}[t]{.32\textwidth}%
		\centering
		\includegraphics[angle=-90, scale=0.21,trim={5cm 0cm 2.5cm 2.5cm},clip ]{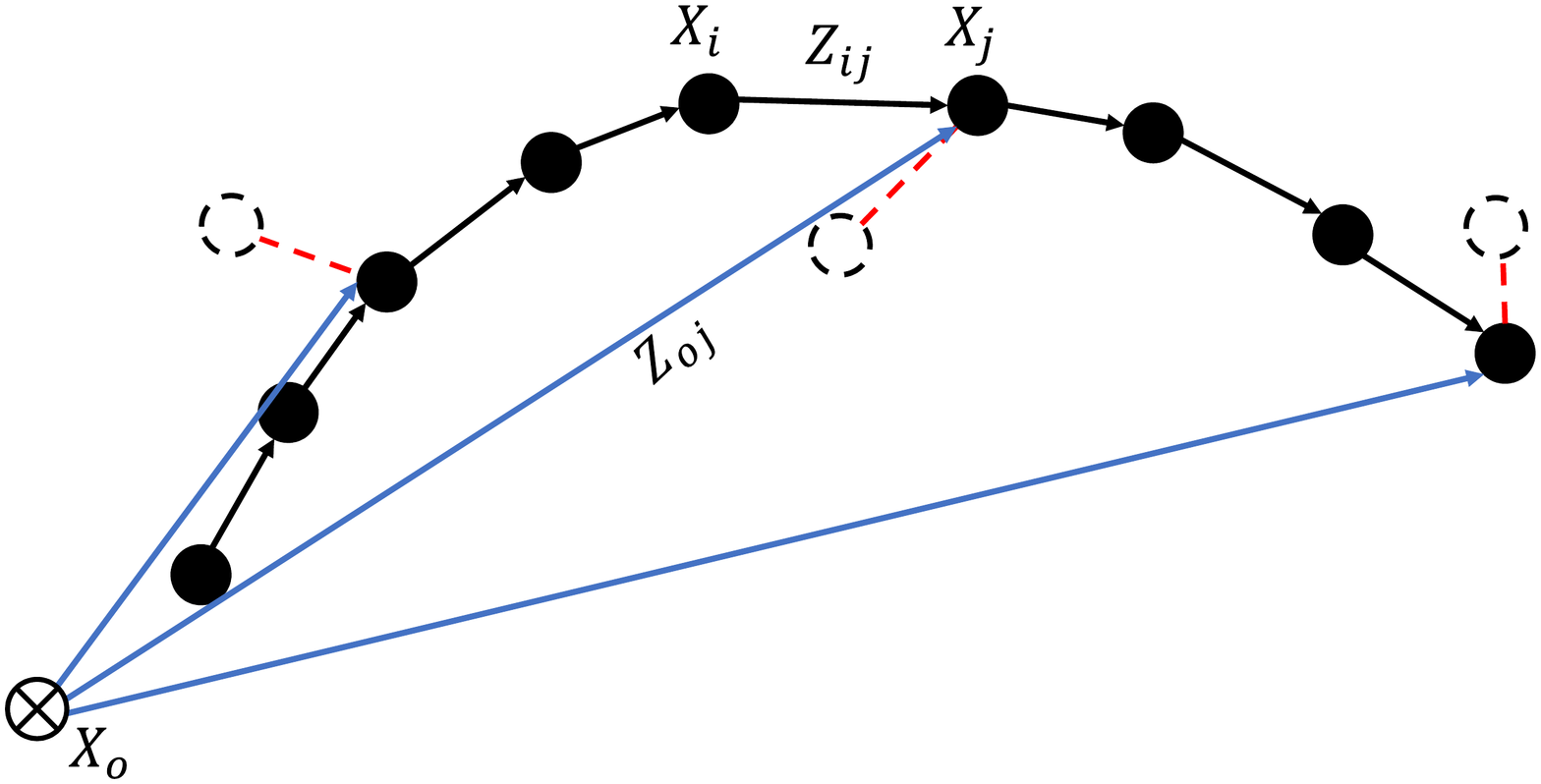}
		\caption{  }
		
		\label{fig:graph1}
	\end{subfigure}	
	\begin{subfigure}[t]{.32\textwidth}%
		\centering
		\includegraphics[angle=-90, scale=0.21,trim={5cm 0cm 2.5cm 2.5cm},clip ]{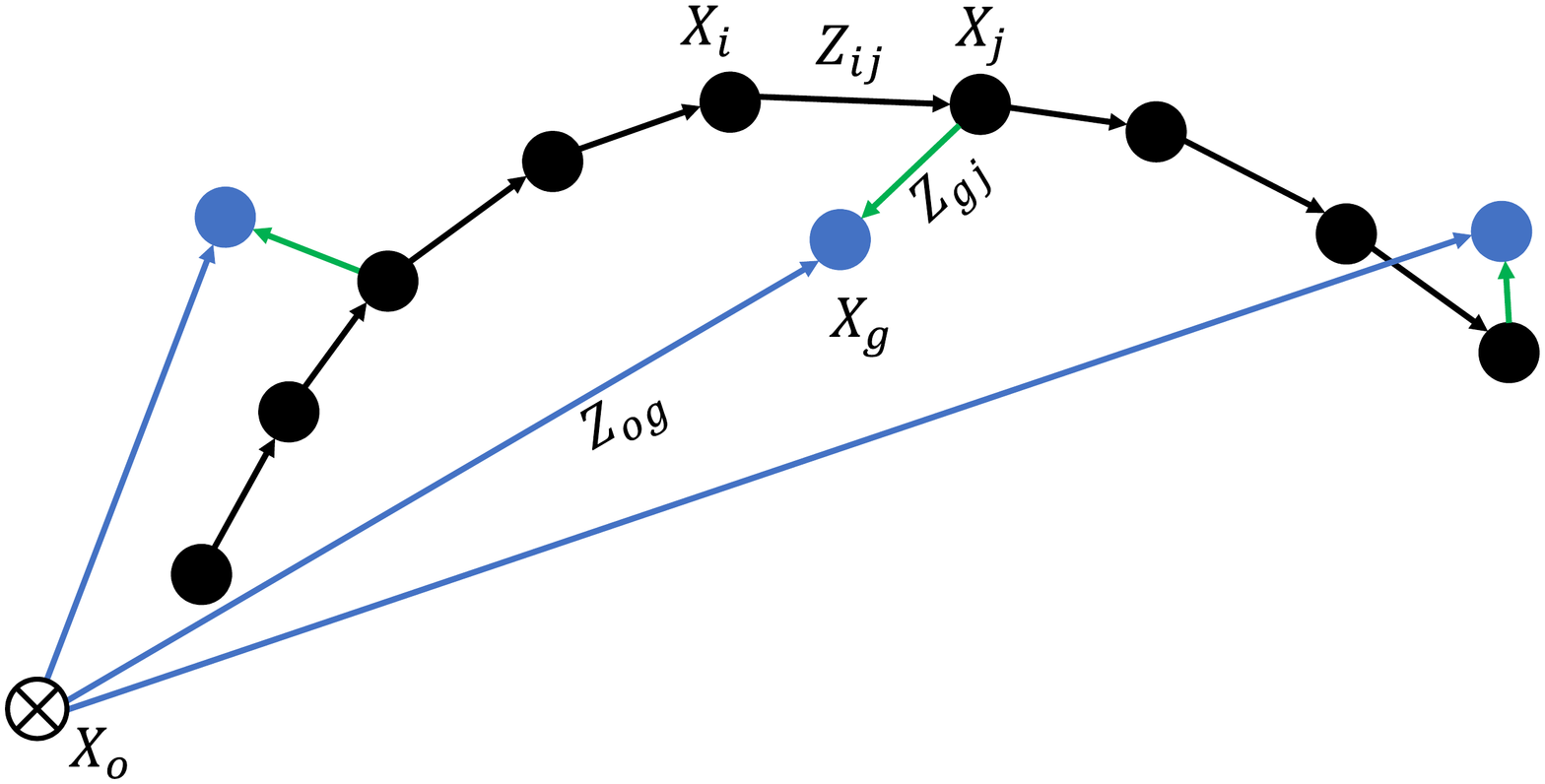}
		\caption{ }
		
		\label{fig:graph2}
	\end{subfigure}
	\begin{subfigure}[t]{.32\textwidth}%
		\centering
		\includegraphics[angle=-90, scale=0.21,trim={5cm 0cm 2.5cm 2.5cm},clip ]{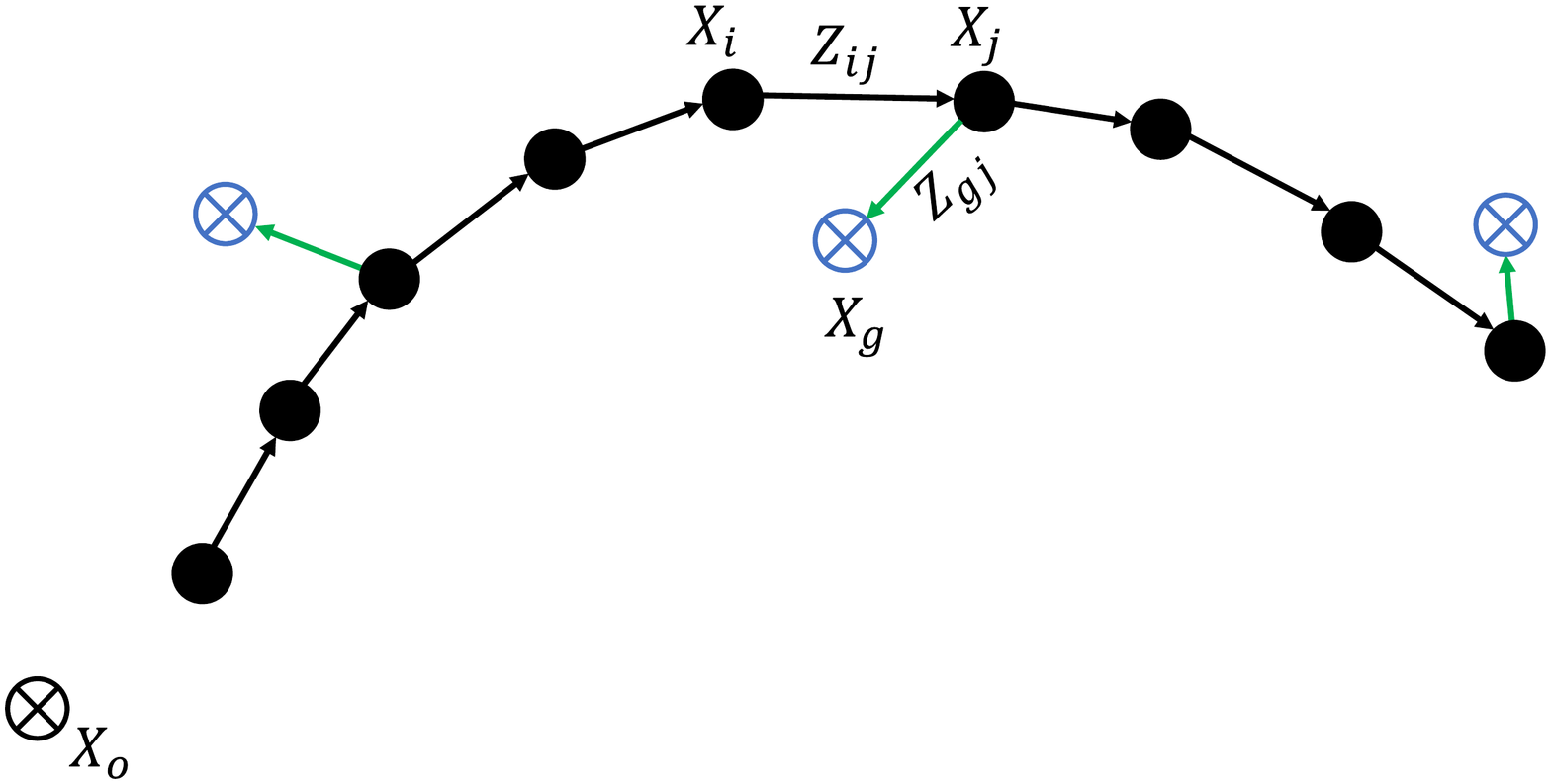}
		\caption{ }
		
		\label{fig:graph3}
		
	\end{subfigure}
	
	\caption{Graph modeling strategies $ G1 $, $ G2 $ and $ G3 $, in (a), (b) and (c) respectively. The black circles are the absolute vehicle poses initialized from the odometry and the black arrows are the corresponding edges. The blue arrows are the GNSS edges connecting the UTM origin node (black circle with a cross) with the corresponding nodes. In (a), the black dashed circles represent the GNSS readings and the error is depicted by red dashed line. In (b), the blue circles are the GNSS readings nodes. The green arrows represent the (virtual) identity edges. In (c), The blue circles with crosses are the GNSS nodes which are kept fixed during optimization.}
	\label{fig:graph}
	
\end{figure*}

%
%
%
%
%
%

\subsection{Graph Structure}\label{sec:graph_struc_gen}

We explore three different strategies to model the optimizable pose-graph: $ G1 $, $ G2 $ and $ G3 $, as shown in Fig. \ref{fig:graph}. The three approaches differ in the manner in which the GNSS readings are modeled. We note that all the three approaches are intrinsically the same, i.e. the global minima of their objective functions is located at the same point in the parameter space. However, due to modeling the GNSS readings differently, their convergence characteristics can differ. 

\textbf{Modeling approach \textit{G1:}} In the first approach, both vehicle odometry and GNSS readings are modeled as measurements (edges), see Fig. \ref{fig:graph1}. The absolute poses of the vehicle are computed from the odometry, hence they initially exactly coincide with the measured relative poses. The goal of the graph optimization is then to minimize the errors related to the GNSS readings. This will alter the relative poses between nodes and hence introduce error, but this error is compensated for by the reduction in the error related to the GNSS readings. After convergence, the graph is in an optimal balance between the errors related to the relative vehicle odometry and the errors related to the absolute GNSS readings.

\textbf{Modeling approach \textit{G2:}} The second approach differs from the first approach because the GNSS readings are now modeled as nodes and their absolute positions are optimized for, see Fig. \ref{fig:graph2}. In order to link the poses of the vehicle, initially provided by the vehicle odometry, to the GNSS readings, we introduce an extra edge between the GNSS nodes and their corresponding vehicle pose, these are depicted by the green arrows in Fig. \ref{fig:graph2}. These edges models (virtual) identity measurements, stating that the particular GNSS poses and the vehicle pose are the same: They act as very strong soft-constraints. The potential benefit of this approach is that there is more flexibility in the graph, which can improve convergence when the vehicle poses are initialized far from a satisfactory local or the global minima. However, by modeling the GNSS readings as nodes, we increase the number of parameters that are optimized for and therefore increase the computational load.

\textbf{Modeling approach \textit{G3:}} In the third approach depicted in Fig. \ref{fig:graph3}, the GNSS readings are also modeled as nodes but now they are kept fixed during optimization. The uncertainties of the GNSS readings are now transferred to the (virtual) identity edges, which  no longer acts as strong soft-constraints but as regular edges. Compared to approach \textit{G1}, this offers an alternative way of modeling the GNSS readings without optimizing for their position as in approach \textit{G2}.

\subsection{Odometry from Velocity and Yaw-rate Sensor}The odometry is generated from the yaw-rate and velocity sensors in the vehicle. Let's consider that the vehicle is traveling from the point $ i $ to $ j $. The change in yaw angle $ a_{ij} $ and displacement $ d_{ij} $ is computed by pre-integrating the yaw rate and velocity at the specific time interval $ t_{ij} $, respectively \cite{imu2015}. The odometry of the vehicle $ o_{ij} $, can be estimated by multiplying the rotation matrix  with the translation vector derived from $ a_{ij} $ and  $ d_{ij} $ respectively. The absolute pose of the vehicle having $ n $ computed odometry measurements, can be estimated by pre-integrating all vehicle odometry measurements upto $ n $. The vehicle odometry drifts at an average 1.1\% of the total driving distance, which is relatively accurate for regular odometry \cite{Ge2012}. 

\subsection{Information Matrix Determination} \label{sec:inf_mat}

In the paper \cite{merf1}, it is assumed that the GNSS sources have constant uncertainties, which is a sub-optimal approach. The GNSS receivers estimate the expected accuracy of their fix for lattitude, lattitude, and altitude at the $ 95\% $ confidence bound. These expected accuracies are provided in meters and denoted with $ epx $, $ epy $, $ epv $. These uncertainty values are computed from GNSS reading Dilution of Precision (DOP). The information matrix values for the UTM-X  and UTM-Y coordinate measurements for each GNSS edge are computed as $ (epx/2)^{-2} $ and  $ (epy/2)^{-2} $ respectively. 

The information matrix for all odometry edges are computed as the inverse of the covariance matrix of each odometry measurement. The covariance matrix is derived from the average 1.1\% drift of the total distance traveled. We assume that when the velocity is zero,  the vehicle cannot move or rotate from its position. In this case, we give the odometry a high certainty ($ 1e^5 $ for the corresponding elements in information matrix) for the corresponding edges. This prevents the vehicle from having abnormal movements after pose-graph optimization, for example the vehicle will not show any lateral displacement with zero longitudinal displacement. 

\subsection{Adaptive GNSS Outlier Rejection} \label{sec:adap_gnss}So far, we have assumed that the $ epx $, $ epy $, $ epv $ values provided by the GNSS receiver are a good estimate for its uncertainty. However, this only holds true in conditions when there is sufficient satellite visibility. In many cases, when the line-of-sight to satellites is blocked or multi-path is induced due to signal reflection, the accuracy of GNSS is severely degraded but the $ epx $, $ epy $, $ epv $ values does not reflect this, i.e. they are highly over-confident. Ignoring this will severely degrade the performance of fusion. We call these erroneous GNSS measurements with over-confident $ epx $, $ epy $, $ epv $ values \textit{outliers} and we propose an approach to detect and ignore them before fusing. This approach is based on the fact that, GNSS readings have low short-term accuracy but the yaw-rate and velocity sensors in the vehicle have very high short-term accuracy, thus the vehicle odometry can be used as an observer to detect GNSS outliers. We do this by computing relative measurements form the absolute GNSS readings for each second and compute the error w.r.t. the relative measurements of the vehicle odometry for the same time span. If the error of change in heading, and displacement is below $ 1.5\deg $ and $ 15m $ respectively, the GNSS reading is considered as a good measurement. We have tuned these thresholds for high precision, i.e. we try to make sure that all outliers are rejected at the expense of also rejecting good measurements. The percentage of GNSS readings rejected is provided in Table \ref{table:dataset_overview}.

\section{Experiments and Results} \label{sec:exp_result}

\begin{figure}[]
	\includegraphics[angle = -90,scale=.29,trim={2.2cm 0cm 2.6cm 0cm},clip]{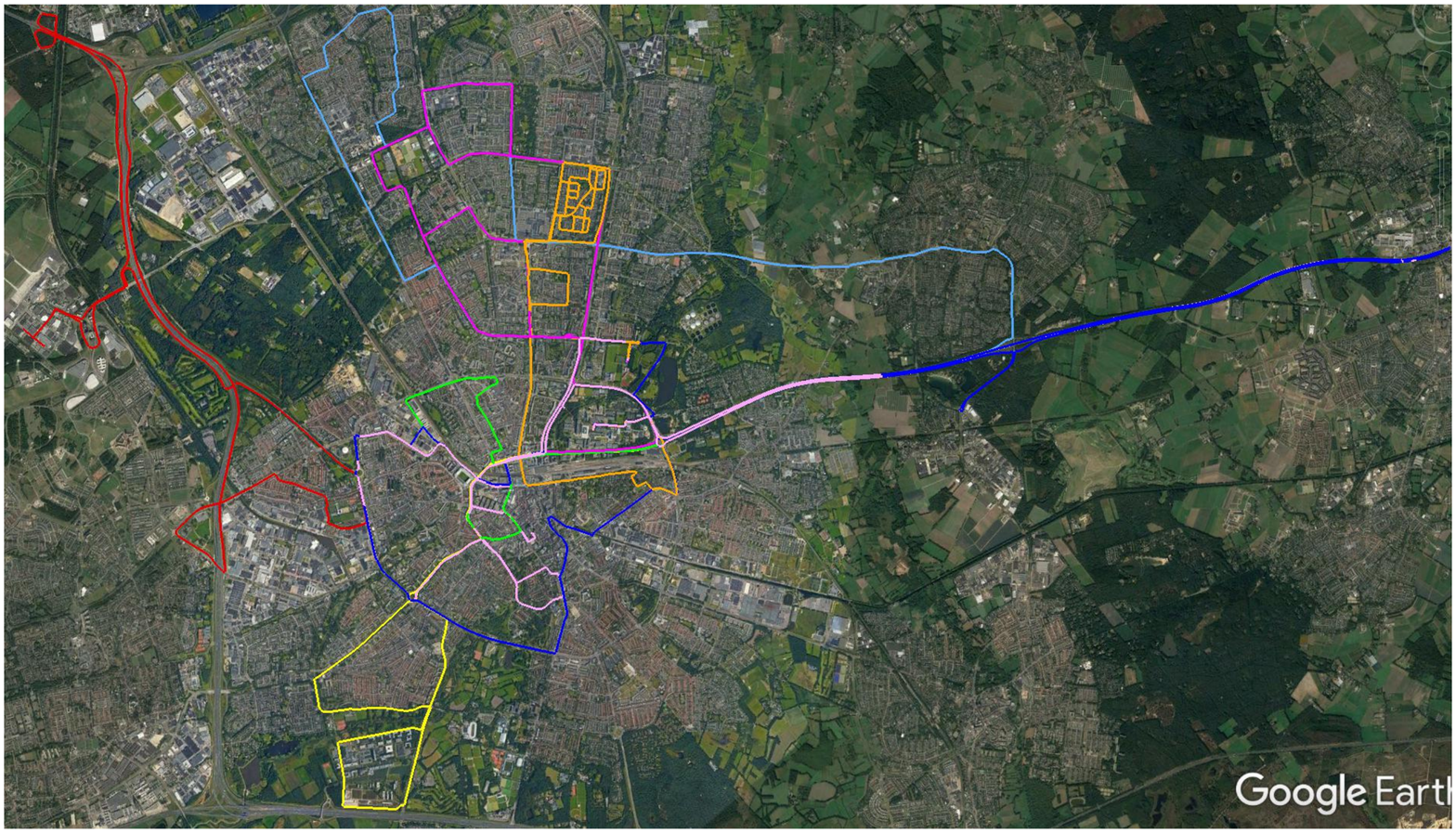}
	\caption{Overview of our 8 datasets. Each dataset is depicted in a different color.}
	\label{fig:dataset_trajectory}
\end{figure}

To evaluate the performance of our sensor fusion approach, we evaluate it on datasets covering more than 180 Km of driving distance. The datasets are recorded with our test vehicle in different environments like urban canyons, highways, towns, etc. in total 8 datasets. The summary of the datasets are provided in the Table \ref{table:dataset_overview} and the trajectories of all datasets are shown in Fig. \ref{fig:dataset_trajectory}. We perform the following experiments for detailed analysis:  

\begin{enumerate}
	\item \textbf{GNSS data analysis:} Evaluation of the quality of GNSS reading form the receiver w.r.t. the RTK-GPS. This sets the baseline for our other experiments.
	\item \textbf{Graph modeling:} We compare and discuss the performance of the three graph modeling approaches provided in Sec. \ref{sec:graph_struc_gen}.
	\item \textbf{Impact of GNSS outlier rejection:} We evaluate the influence of adaptive GNSS outlier rejection, as introduced in Sec. \ref{sec:adap_gnss}.

\end{enumerate}

\begin{table}[]
	\centering
	\caption{Dataset overview}
	\label{table:dataset_overview}
	\begin{tabular}{C{.8cm}C{.8cm}C{1cm}C{1cm}C{1.1cm}C{1.4cm}}
		Dataset & Distance (km) & \#GNSS points & \#Odometry points & GNSS readings rejected & Environment \\ \hline
		
		1       & 24         & 1814              & 46131	&14.88\% & Highway                 \\
		2       & 8          & 2495              & 63042   &26.18\% & Urban              \\
		3       & 37         & 3354              & 85226   &57.43\% & Highway/rural              \\
		
		4       & 25         & 2716              & 68351   &24.30\% & Rural             \\
		5       & 19         & 2458              & 61479   &24.41\% & Urban             \\
		6       & 24         & 1585              & 39685   &25.91\% & Highway             \\
		7       & 26         & 3281              & 82313   &9.53\% & Rural            \\
		
		8       & 25         & 4188              & 104949  &29.37\% & Rural            \\ \hline
	\end{tabular}
\end{table}

\subsection{Vehicle Setup} Our experimental vehicle is equipped with a U-Blox GNSS receiver with PPP, a RTK-GPS system, and a CAN interface to access the vehicle ECU messages. Post-processed RTK-GPS is the state-of-the-art industry standard for accurate GNSS based localization, which has an accuracy of up-to $ 0.01m $. The yaw-rate in rad/s from the yaw-rate sensor and velocity in m/s from the velocity sensor  is received at 25 Hz through the CAN interface. The U-Blox GNSS receiver operates at 1 Hz. The clock of the computer in the vehicle is kept synchronized with the global GNSS satellite clock using the Pulse-per-second (PPS) from the GNSS receiver.    


\subsection{Performance Metrics} To evaluate the performance of our sensor fusion algorithm, we compare the results with the  post-processed RTK-GPS using the metrics described below. For both the GNSS readings and the fusion results, we compute these metrics, for the poses corresponding to the PPS, which we call \textit{PPS-Poses}.
\begin{enumerate}
\item \textbf{Maximum offset error} ($ Max. $) in $ meters $, which is the maximum offset euclidean distance error over all \textit{PPS-Poses} of each dataset, computed as
\begin{equation}
Max.=\max_{1 \leq i \leq n}\sqrt{(\bar{X}_{i}-\hat{X}_{i})^2 + (\bar{Y}_{i}-\hat{Y}_{i})^2}
\end{equation}
, where $ n $ is the total number of poses in a dataset, $ \bar{X} $, $ \bar{Y} $ and $ \hat{X} $, $ \hat{Y} $ are the corresponding co-ordinate values of the localization system GNSS or GNSS-Odometry fusion, and of the RTK-GPS points respectively.
\item \textbf{Accuracy} ($ Acc. $) in $ meters $, which is the euclidean distance of the point computed from the mean offset error in UTM-X and UTM-Y axes of the UTM co-ordinate system for each dataset. The accuracy represents the structural offset between the RTK-GPS and the localization system under consideration. It is computed as 
\begin{equation}
Acc.=\sqrt{\mu_{X}^2+\mu_{Y}^2}
\end{equation}
, where $ \mu_{X} $ and $ \mu_{Y} $ is the mean offset in the considered localization system, computed as
\begin{equation}
\label{eq:ux}
\mu_{X} = \dfrac{1}{n}\sum_{i=1}^{n}\bar{X}_{i}-\hat{X}_{i}
\end{equation}
\begin{equation}
\label{eq:uy}
\mu_{Y} = \dfrac{1}{n}\sum_{i=1}^{n}\bar{Y}_{i}-\hat{Y}_{i}.
\end{equation}
\item \textbf{Precision} ($ Prec. $) in $ meters $, which is the standard deviation of the distance of each point from the computed mean offset error for UTM-X and UTM-Y axes for each dataset. It represents the variation or dispersion of the readings for the considered localization system from its mean for each dataset. It is computed as
\begin{equation}
Prec.=\sqrt{\dfrac{1}{n-1}\sum_{i=1}^{n}D_i^2}
\end{equation}
, where $ D_i $ is the distance of each point from the mean off-set, i.e.
\begin{equation}
D_i = \sqrt{(\bar{X}_{i} - \mu_{X})^2+(\bar{Y}_{i} - \mu_{Y})^2}      .
\end{equation}

\end{enumerate}

We also report the averages over all dataset and the relative percentage improvements of the pose-graph fusion with respect to the GNSS.

%
%

\subsection{Results}
\subsubsection{\textbf{GNSS data analysis}} \label{exp:exp1} 
\begin{figure}[!ht]
	\centering
	\begin{subfigure}[t]{.11\textwidth}%
		\centering
		\includegraphics[trim={3.4cm 2.5cm 2.3cm 1.8cm},clip,width=1.75cm,height=1.755cm]{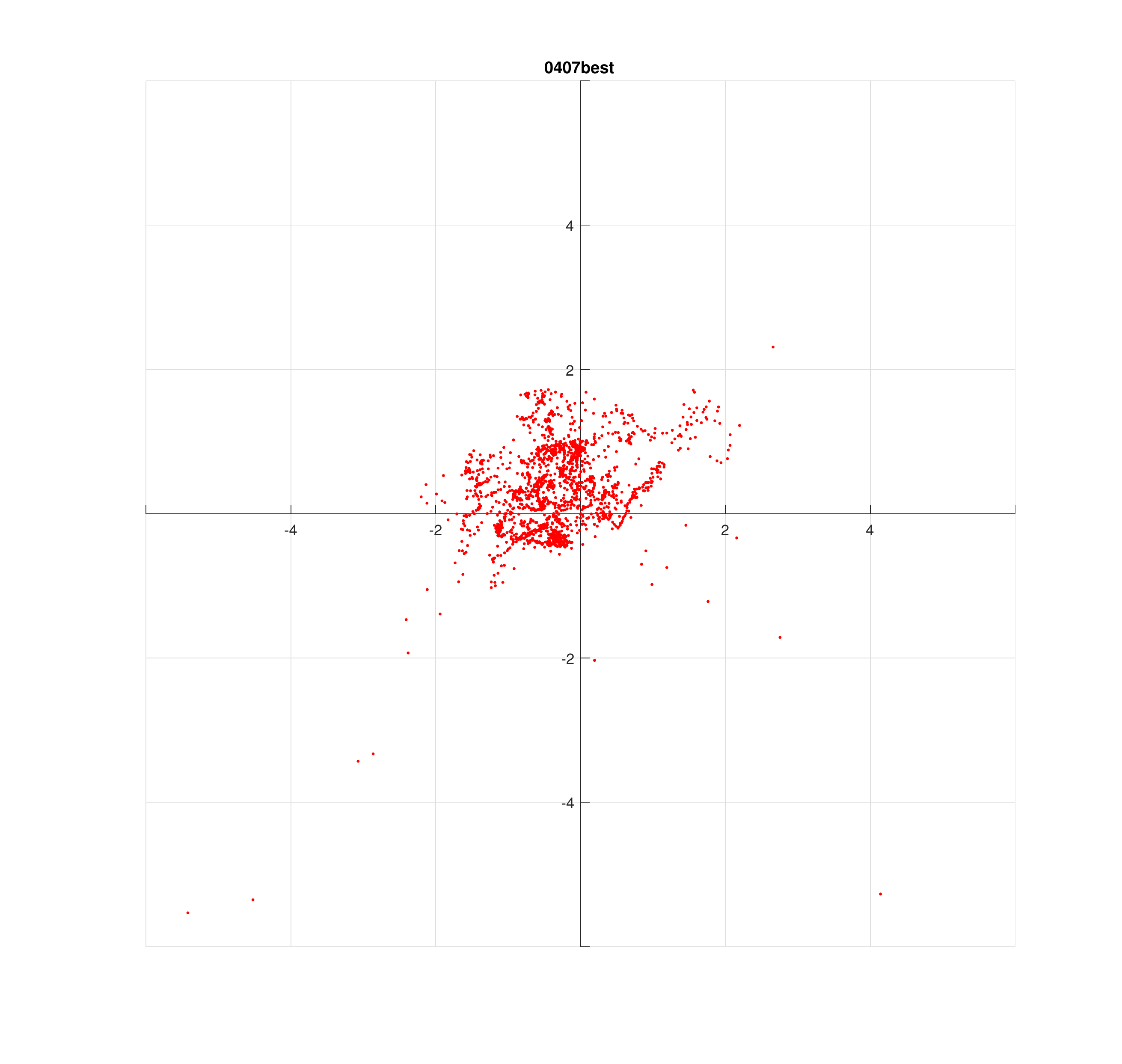}
		\caption{}
		\label{fig:datset1}
	\end{subfigure}
	\begin{subfigure}[t]{.11\textwidth}%
		\centering
		\includegraphics[trim={3.4cm 2.5cm 2.3cm 1.8cm},clip,width=1.75cm,height=1.75cm ]{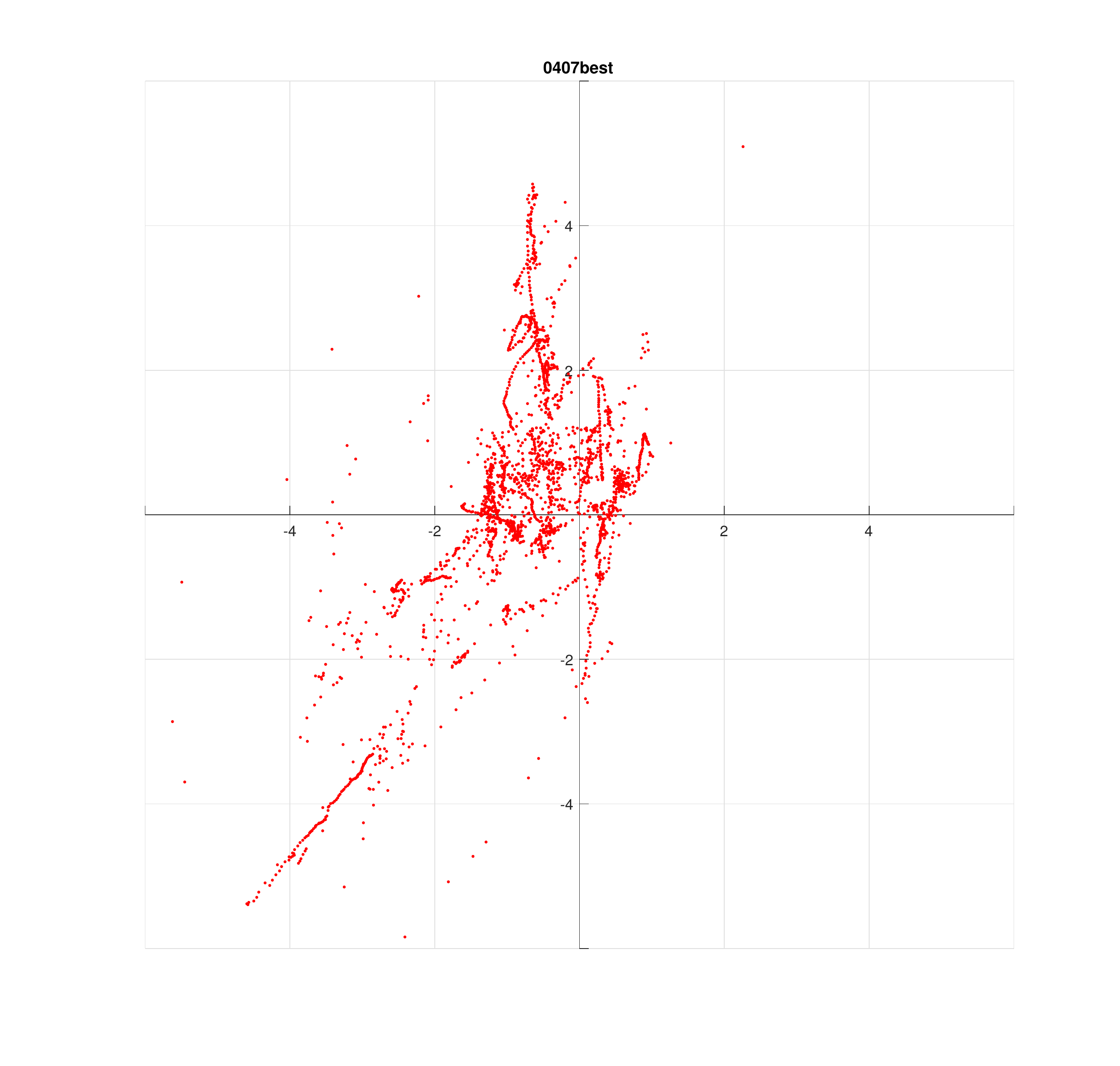}
		\caption{}
		\label{fig:datset2}
	\end{subfigure}
	\begin{subfigure}[t]{.11\textwidth}%
		\centering
		\includegraphics[trim={3.4cm 2.5cm 2.3cm 1.8cm},clip,width=1.75cm,height=1.75cm ]{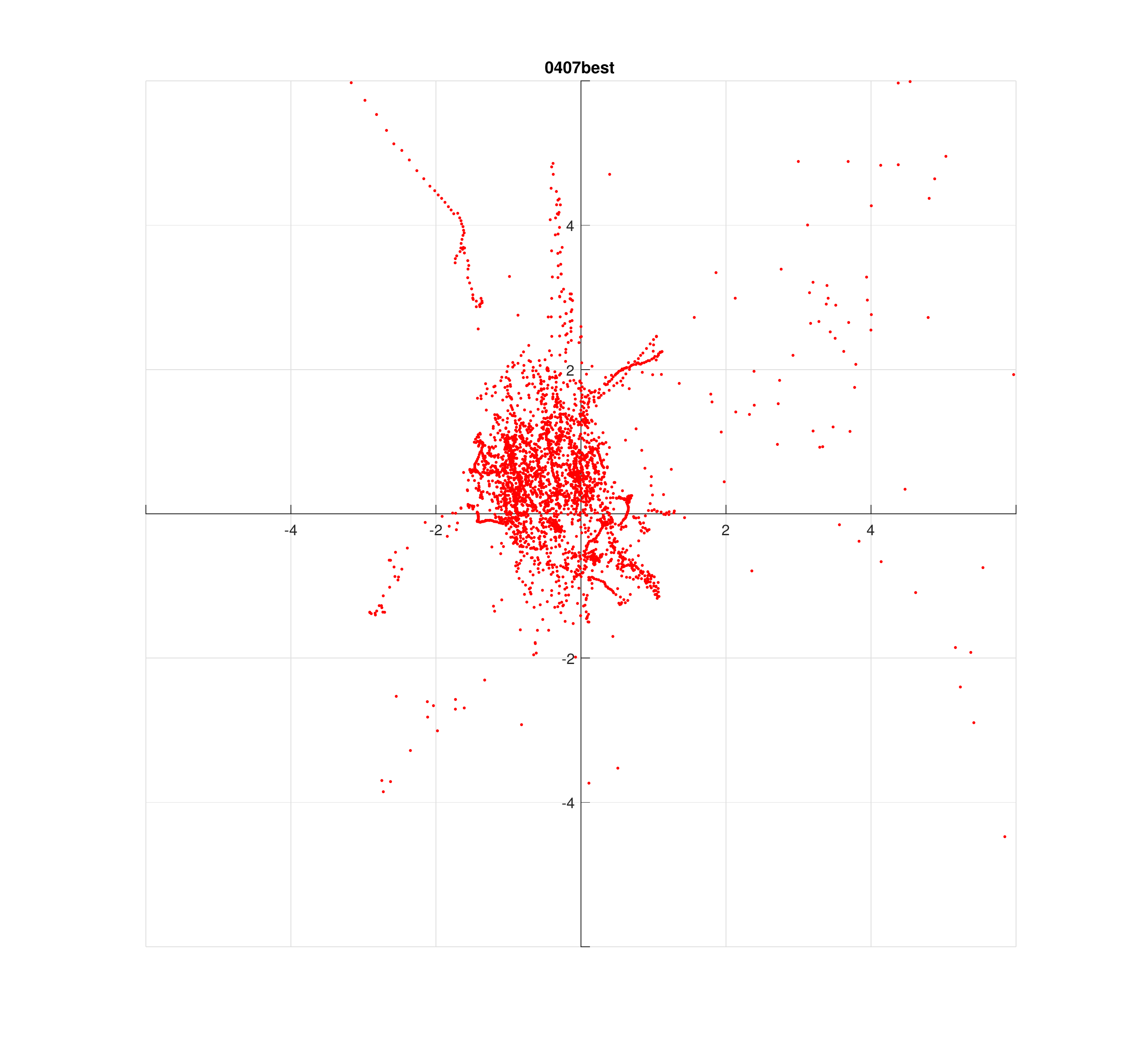}
		\caption{}
		\label{fig:datset3}
	\end{subfigure}
	\begin{subfigure}[t]{.11\textwidth}%
		\centering
		\includegraphics[trim={3.4cm 2.5cm 2.3cm 1.8cm},clip,width=1.75cm,height=1.75cm]{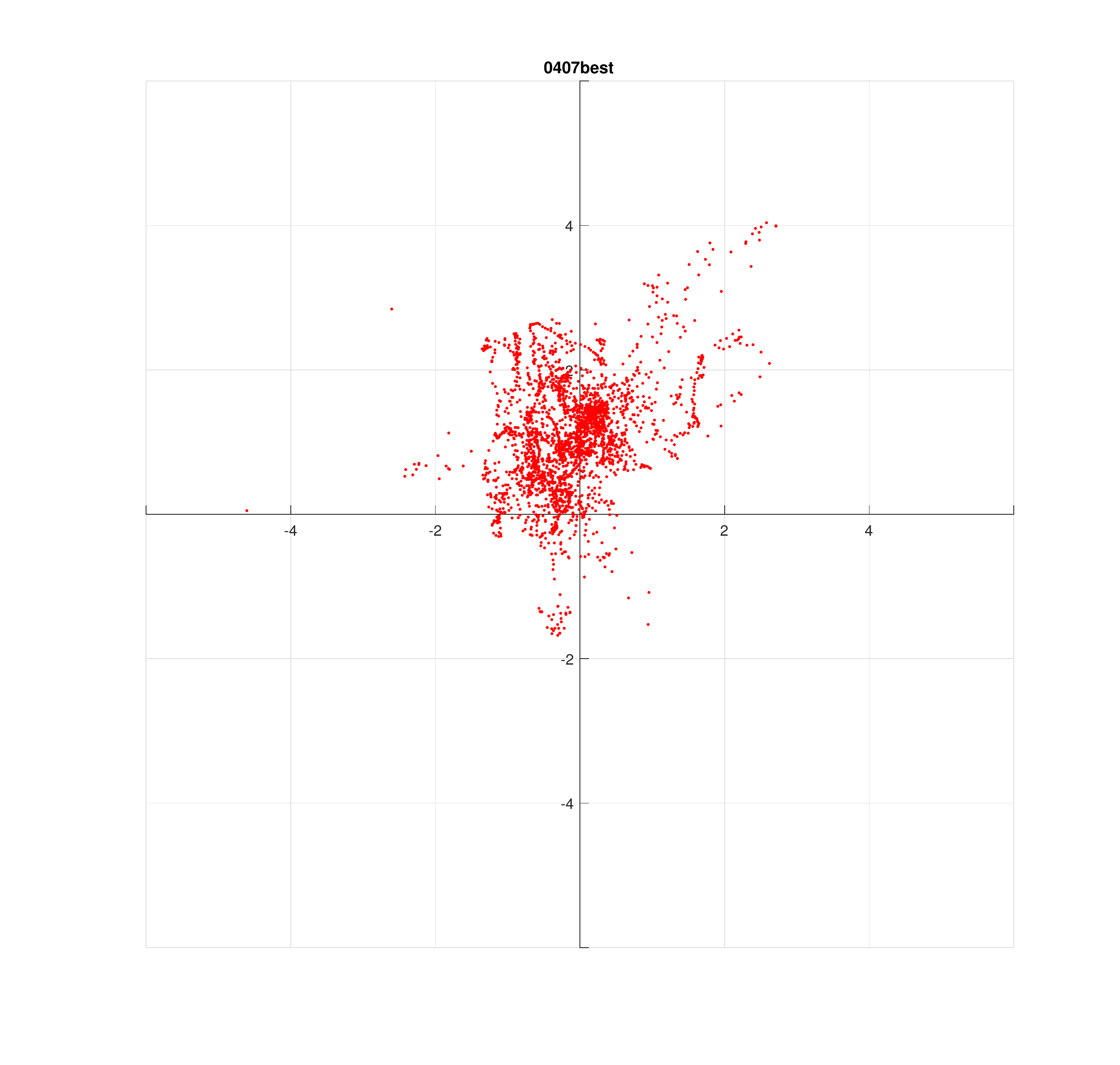}
		\caption{}
		\label{fig:datset4}
	\end{subfigure}

	\begin{subfigure}[t]{.11\textwidth}%
		\centering
		\includegraphics[trim={3.4cm 2.5cm 2.3cm 1.8cm},clip,width=1.75cm,height=1.75cm]{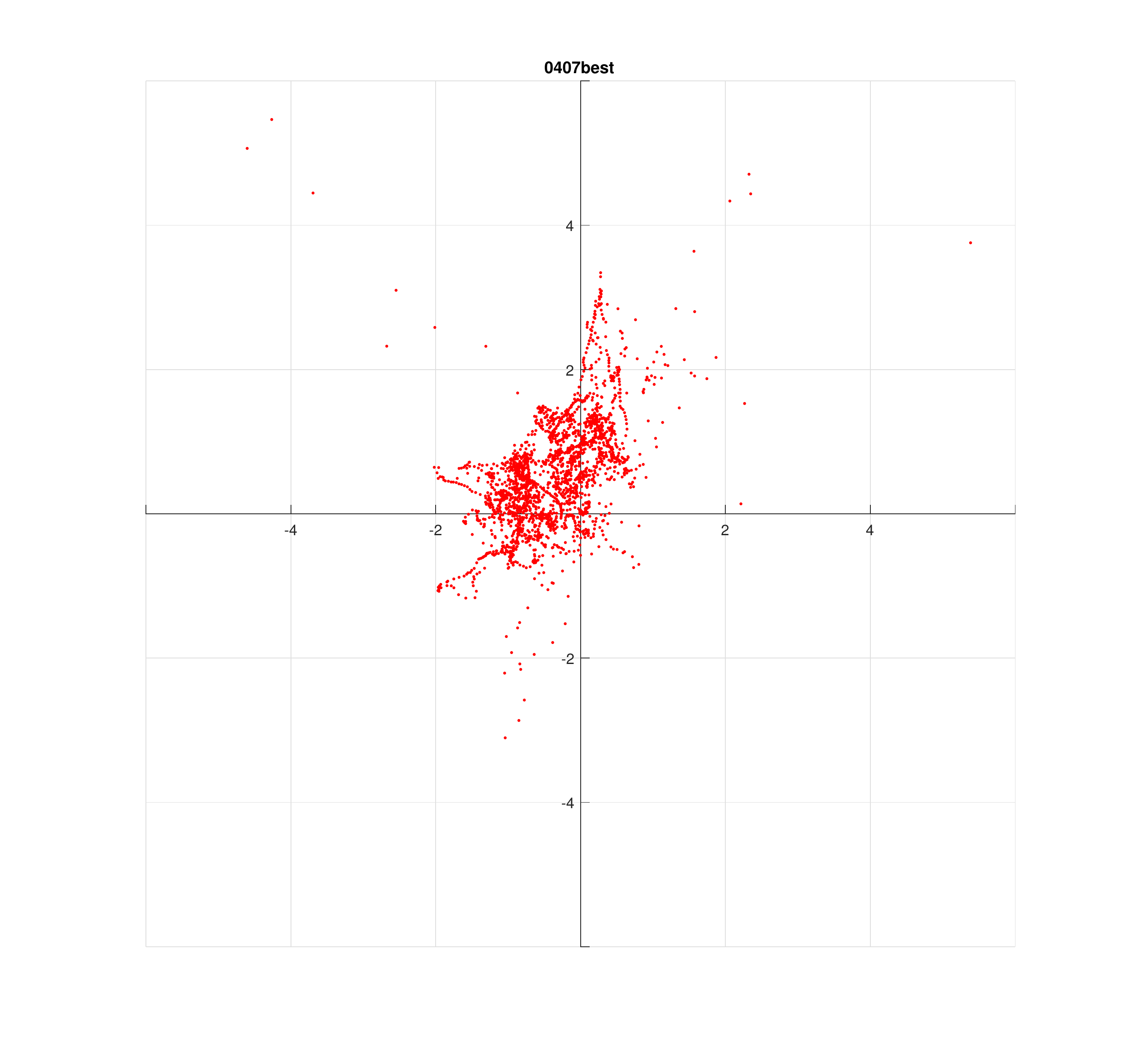}
		\caption{}
		\label{fig:datset5}
	\end{subfigure}
	\begin{subfigure}[t]{.11\textwidth}%
		\centering
		\includegraphics[trim={3.4cm 2.5cm 2.3cm 1.8cm},clip,width=1.75cm,height=1.75cm]{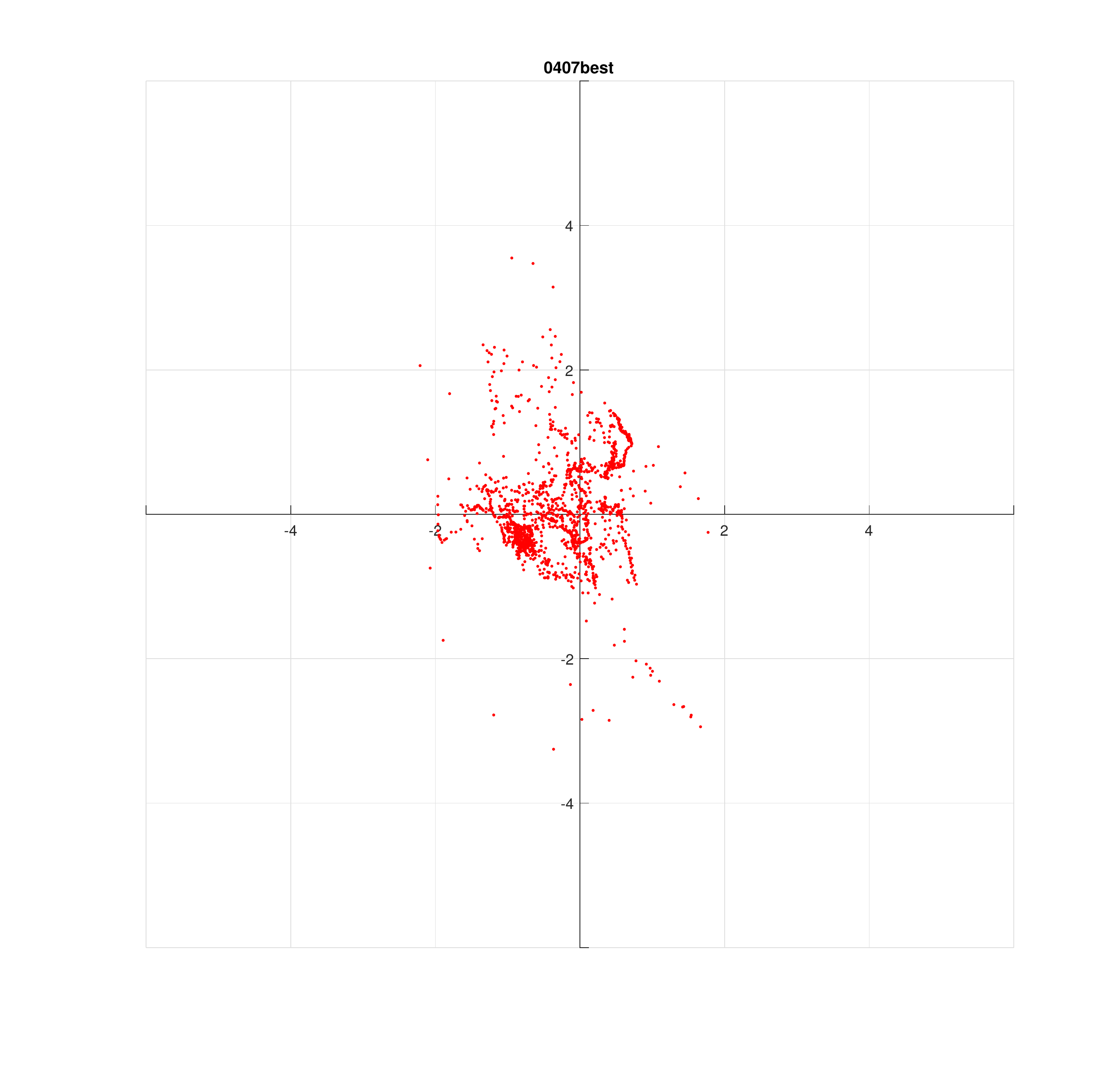}
		\caption{}
		\label{fig:datset6}
	\end{subfigure}
	\begin{subfigure}[t]{.11\textwidth}%
		\centering
		\includegraphics[trim={3.4cm 2.5cm 2.3cm 1.8cm},clip ,width=1.75cm,height=1.75cm]{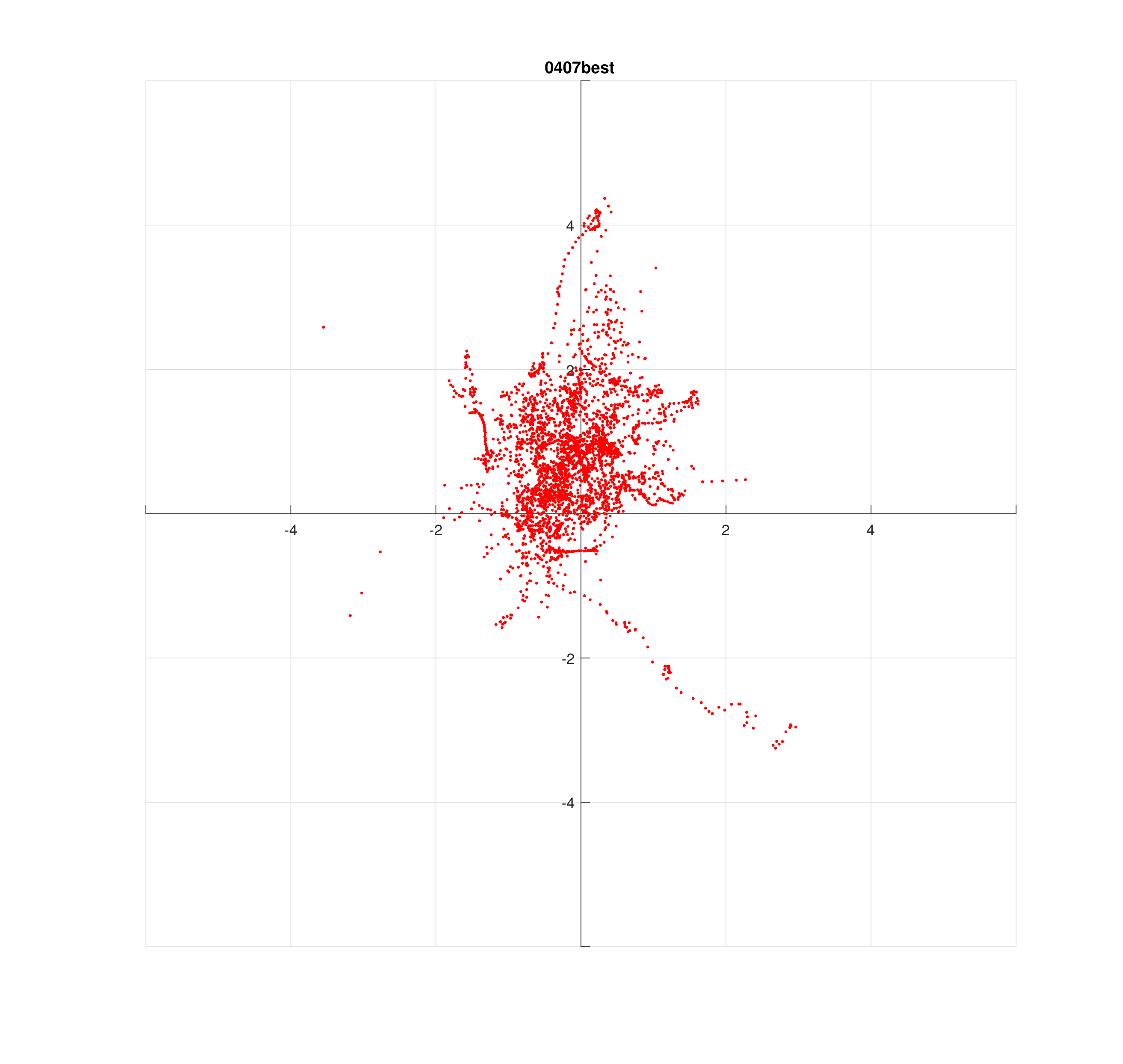}
		\caption{}
		\label{fig:datset7}
	\end{subfigure}
	\begin{subfigure}[t]{.11\textwidth}%
		\centering
		\includegraphics[trim={3.4cm 2.5cm 2.3cm 1.8cm},clip,width=1.75cm,height=1.75cm]{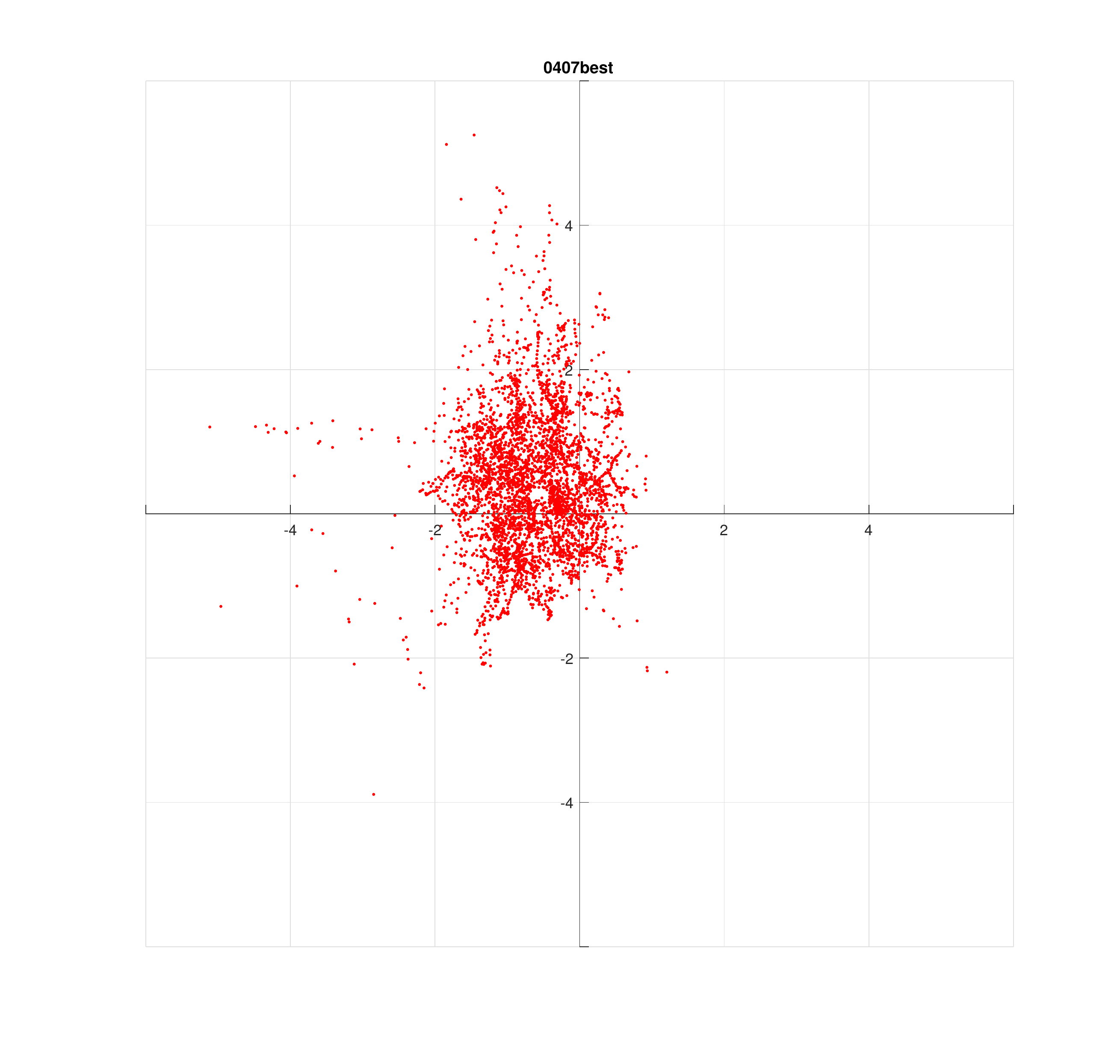}
		\caption{}
		\label{fig:datset8}
	\end{subfigure}
	
	\caption{GNNS positions wrt. the RTK-GPS ground truth for the eight datasets. The plots with tile size of 2x2 $ m $, clearly show that GNSS errors are biased, and correlated, causing non-zero mean behavior for time spans up to 65 minutes. }
	\label{fig:GNSS_offset}
\end{figure}

First, we evaluate the quality of the GNSS data from the receiver w.r.t. the RTK-GPS for all datasets. The performance metrics for the GNSS readings are shown in Table \ref{table:GNSS_metrics}. The accuracy of all the datasets are never close to zero, due to an error bias in the GNSS readings. This bias is clearly visible in the error scatter plots depicted in Fig. \ref{fig:GNSS_offset}. It is evident that over time spans that are relevant for automotive, i.e. minutes to hours, the GNSS error distribution does not exhibit zero-mean behavior and that GNSS errors are highly correlated in time. Please note that the existence of this bias is exactly the reason why RTK-GPS base stations need at-least 24 hours or more averaging time to achieve a localization accuracy of 2 cm. Fusing odometry measurements with GNSS readings can improve the precision of the localization system but it cannot remove this bias and therefore it also cannot improve accuracy, as the vehicle odometry only provides relative localization information \cite{introgps}.

\begin{table}[!ht]
	\centering
	\caption{Performance metrics of GNSS receiver}
	\label{table:GNSS_metrics}
	\begin{tabular}{cccc}
		Dataset & $ Max. $    & $ Acc. $    & $ Prec. $ \\ \hline
		1       &66.291	    &0.489	&2.653 \\
		2       & 21.799	&0.874	&2.491 \\
		3       & 49.068	&0.715	&2.293 \\
		4       & 4.828	    &1.089	&1.025 \\
		5       & 16.776	&0.678	&1.141 \\
		6       & 3.672	    &0.336	&0.954 \\
		7       & 7.775	    &0.782	&1.177 \\
		8       & 18.037	&0.805	&1.264 \\ \hline\hline
		\textbf{Average} & \textbf{23.531} & \textbf{0.721} & \textbf{1.625} \\ \hline
	\end{tabular}
\end{table}


\begin{figure*}[!ht]
	\centering
	\begin{subfigure}[t]{.32\textwidth}%
		\centering
		\includegraphics[scale=.165,trim={2cm 1cm 2cm 1cm},clip]{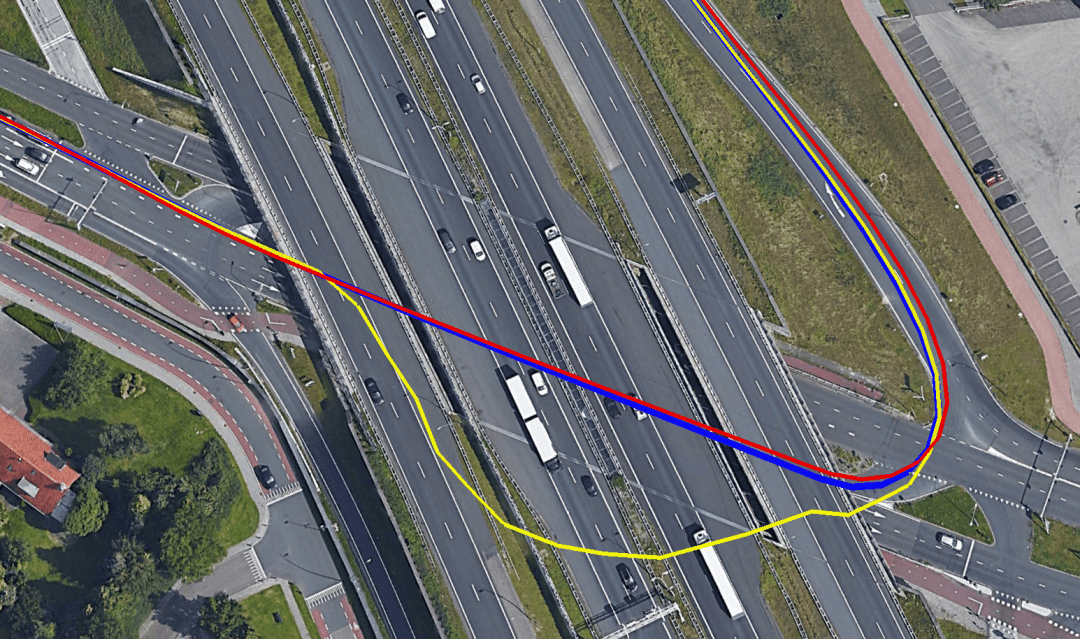}
		\caption{}
		\label{fig:under_br1}
	\end{subfigure}
	\begin{subfigure}[t]{.32\textwidth}%
		\centering
		\includegraphics[scale=.165,trim={2cm 1cm 2cm 1cm},clip]{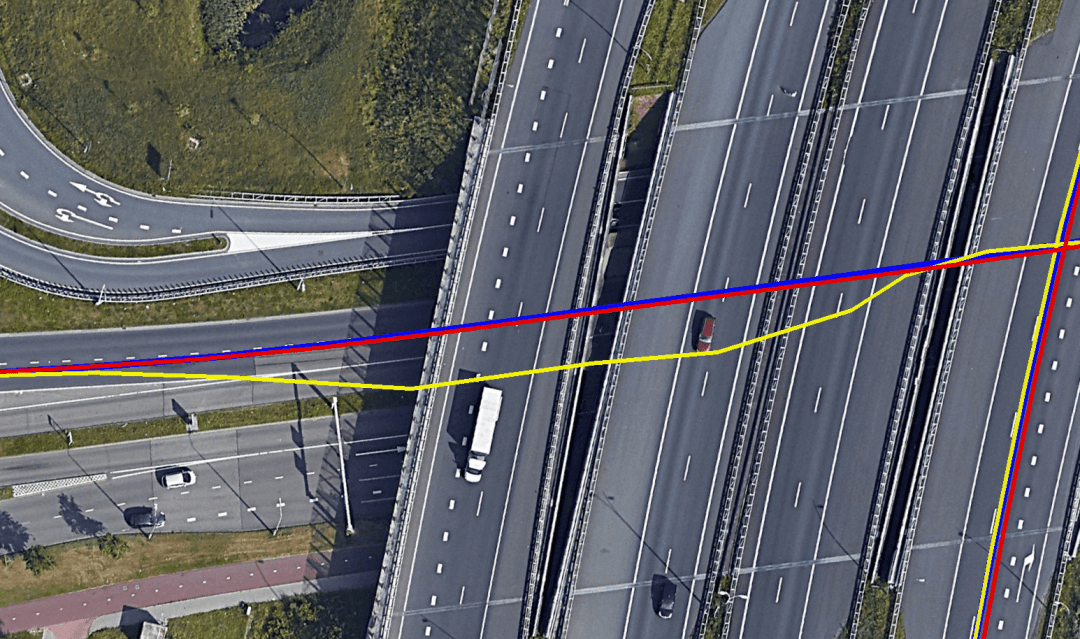}
		\caption{}
		\label{fig:under_br2}
	\end{subfigure}
	\begin{subfigure}[t]{.32\textwidth}%
		\centering
		\includegraphics[scale=.165,trim={2cm 1cm 2cm 1cm},clip]{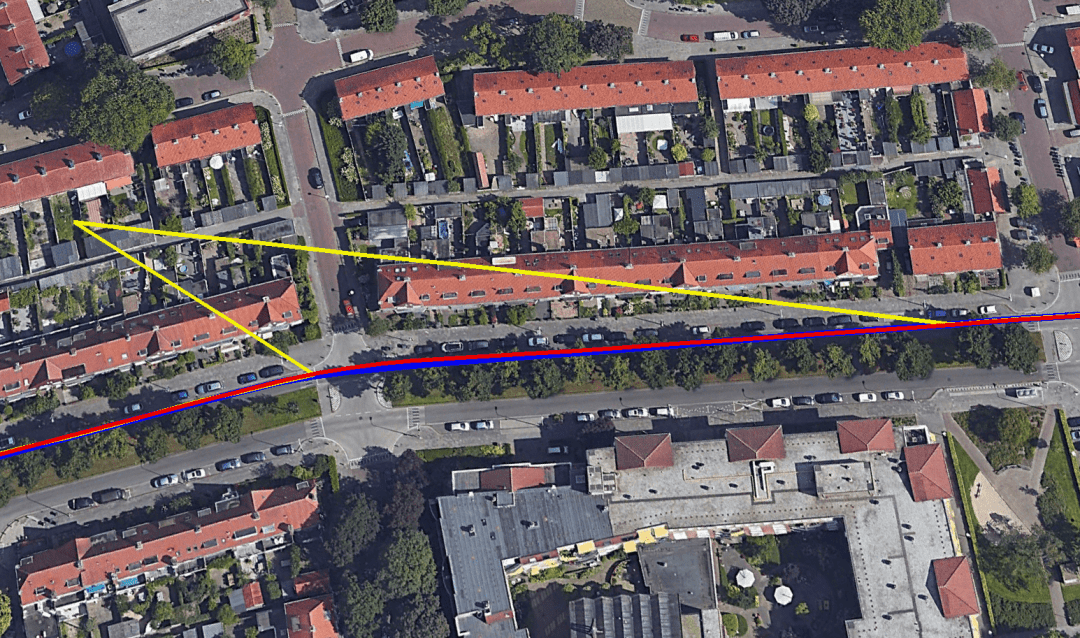}
		\caption{}
		\label{fig:trees1}
	\end{subfigure}
	\caption{Fusion performance of graph modeling approach \textit{G2}: RTK-GPS (Red), GNSS (Yellow), Fusion (Blue)  }
	\label{fig:fusion_results}
\end{figure*}

\subsubsection{\textbf{Graph modeling}} \label{sec:graph_modeling_approaches}

 The performance metrics of the modeling approach G\textit{1}, is shown in the Table \ref{table:graph1}. The  average \textit{Max.} decreased by $ 21.84\% $ but the average \textit{Prec.} increased by $ 10.11\% $. The average \textit{Acc.} remained close to the average GNSS accuracy, as expected. It is observed that out of the 8 experiments only 3 converged. We believe this convergence issue is caused by a too rigid modeling of the pose-graph, which hampers convergence when the vehicle poses, initialized form the odometry, are far off from the GNSS readings. 

The pose-graph modeling approach \textit{G2} performed much better than \textit{G1}. The optimization converged for all the 8 experiments. Table \ref{table:graph2} shows the performance metrics for this approach. The average \textit{Max.} and \textit{Prec.} decreased by $ 69.53\% $ and $ 17.79\% $ respectively. It performed well in both highways and urban-area scenarios. Fig. \ref{fig:fusion_results} shows some scenarios where the fusion algorithm clearly shows benefit over non-fused GNSS. We believe that the improvement with respect to the first model \textit{G1}, is due to the fact that model \textit{G2} has more flexibility, as the GNSS positions are also optimized for, which improves convergence for challenging initialization.  



The pose-graph modeling approach \textit{G3} performed well but the performance was less than \textit{G2}. Table \ref{table:graph3} shows the performance metrics of this approach. The average \textit{Max.} and \textit{Prec.} decreased by $ 49.50\% $ and $ 15.33\% $ respectively. Like \textit{G1} and \textit{G2} there is no significant change in the average accuracy, as expected. 

\begin{table}[!ht]
	\centering
	\caption{Performance metrics table for Graph model \textit{G1}}
	\label{table:graph1}
	\begin{tabular}{cccc}
		Dataset                      & $ Max. $    & $ Acc. $    & $ Prec. $  \\ \hline
		1                            & 9.568	&0.451	&1.028		   \\
		2                            & 20.835	&0.882	&2.550		   \\
		3                            & 28.505	&0.709	&2.014		   \\
		4                            & 16.242	&1.136	&1.691		   \\
		5                            & 32.260	&0.698	&2.134		   \\
		6                            & 15.498	&0.315	&1.172		   \\
		7                            & 6.171	&0.779	&1.215		   \\
		8                            & 18.049	&0.823	&2.150		   \\ \hline \hline
		\textbf{Average}                      & \textbf{18.391} & \textbf{0.724} & \textbf{1.744}   \\
		\textbf{Improvement w.r.t. GNSS (\%)} & \textbf{21.843} & \textbf{-0.432} & \textbf{-7.364} \\ \hline
	\end{tabular}
\end{table}

\begin{table}[!ht]
	\centering
	\caption{Performance metrics table for Graph model \textit{G2}}
	\label{table:graph2}
	\begin{tabular}{cccc}
		Dataset                      & $ Max. $    & $ Acc. $    & $ Prec. $ \\ \hline
		1                            & 6.282	&0.462	&0.956		  \\
		2                            & 14.173	&0.788	&2.339		  \\
		3                            & 14.015	&0.727	&1.679		  \\
		4                            & 4.846	&1.092	&1.133		  \\
		5                            & 4.266	&0.653	&1.010		  \\
		6                            & 3.117	&0.320	&0.928		  \\
		7                            & 5.297	&0.798	&1.295		  \\
		8                            & 5.367	&0.796	&1.346		  \\ \hline \hline
		\textbf{Average}                      & \textbf{7.170}  & \textbf{0.705} & \textbf{1.336}  \\
		\textbf{Improvement w.r.t. GNSS (\%)} & \textbf{69.528} & \textbf{2.282} & \textbf{17.794} \\ \hline
	\end{tabular}
\end{table}

\begin{table}[!ht]
	\centering
	\caption{Performance metrics table for Graph model \textit{G3}}
	\label{table:graph3}
	\begin{tabular}{cccc}
		Dataset                      & $ Max. $    & $ Acc. $    & $ Prec. $ \\ \hline
		1                            & 10.616	&0.441	&1.040		  \\
		2                            & 20.077	&0.858	&2.378		  \\
		3                            & 27.146	&0.709	&1.975		  \\
		4                            & 5.408	&1.096	&1.050		  \\
		5                            & 6.836	&0.657	&0.986		  \\
		6                            & 3.046	&0.335	&0.931		  \\
		7                            & 15.186	&0.762	&1.365		  \\
		8                            & 6.758	&0.800	&1.280		  \\ \hline \hline
		\textbf{Average}                      & \textbf{11.884} & \textbf{0.707} & \textbf{1.376}  \\ 
		\textbf{Improvement w.r.t. GNSS (\%)} & \textbf{49.495} & \textbf{1.895} & \textbf{15.326}  \\ \hline
	\end{tabular}
\end{table}

\subsubsection{\textbf{Impact of GNSS outlier rejection}} In this experiment, the optimizable pose-graph was generated using approach \textit{G2} but now only by applying \textit{Information Matrix Determination} strategy (Sec. \ref{sec:inf_mat}) and without applying \textit{Adaptive GNSS Outlier Rejection} strategy. The performance metrics of this approach are shown in Table \ref{table:infmat}. The average \textit{Max.} and \textit{Prec.} decreased by $ 54.23\% $ and $ 8.87\% $ respectively. The performance mostly degraded in urban canyon areas, this is because the GNSS uncertainty values used to calculate the information matrix becomes unreliable on low satellites visibility and unwanted reflection of GNSS signals.
It can be concluded that, using \textit{Adaptive GNSS Outlier Rejection} strategy, along with Information matrix computation, clearly improves the performance of the fusion strategy \textit{G2}, with further reduction in both maximum offset error and standard deviation by around $ 15\% $ and $ 9\%  $ respectively, as seen in Tab. \ref{table:graph2}. 

\begin{table}[!ht]
	\centering
	\caption{Performance metrics table without GNSS outlier rejection computation for Graph model \textit{G2}}
	\label{table:infmat}
	\begin{tabular}{cccc}
		Dataset                      & $ Max. $    & $ Acc. $    & $ Prec. $ \\ \hline
		1                            & 17.580	&0.473	&1.628		  \\
		2                            & 15.028	&0.833	&2.293		  \\
		3                            & 29.872	&0.686	&2.469		  \\
		4                            & 4.841	&1.074	&1.072	      \\
		5                            & 4.268	&0.658	&0.973		  \\
		6                            & 3.055	&0.323	&0.926		  \\
		7                            & 5.265	&0.785	&1.226		  \\
		8                            & 6.244	&0.808	&1.256		  \\ \hline \hline
		\textbf{Average}                      & \textbf{10.769} & \textbf{0.705} & \textbf{1.481}  \\
		\textbf{Improvement w.r.t. GNSS (\%)} & \textbf{54.234} & \textbf{2.222} & \textbf{8.873} \\ \hline
	\end{tabular}
\end{table}

\section{CONCLUSIONS} \label{sec:conclusion}


We have developed a pose-graph generation framework for fusing vehicle odometry and GNSS readings, which is validated on 8 datasets covering more than 180 km. The experiments showed a clear reduction in outliers and improvements in precision of the localization system in challenging scenarios. It is shown that, the graph model in which the GNSS readings are modeled as optimizable nodes, achieves best results in our experiments, as it allows for more flexibility and thereby improves convergences. Furthermore, the ability to detect GNSS outliers using vehicle odometry and to replace those outliers with integrated vehicle odemetry, is shown to be key to achieving optimal GNSS-odometry fusion results.

\addtolength{\textheight}{-12cm}   





\bibliographystyle{ieeetran}
\bibliography{References2}

\end{document}